\documentclass[twocolumn,aps,showpacs,preprintnumbers,amsmath,amssymb, superscriptaddress,longbibliography]{revtex4-1}
\usepackage{bm}
\usepackage{graphicx}
\usepackage{color}
\usepackage{xspace}
\definecolor{dblue}{rgb}{0,0,0.6}
\definecolor{dred}{rgb}{0.9,0,0}
\definecolor{dgreen}{rgb}{0,0.4,0}

\pdfoutput=1

\begin{document}

\title{Robustness of topological states to lattice instability in non-symmorphic topological insulator KHgSb}

\author{D.~Chen}\thanks{These authors contributed equally to this work}
\affiliation{Beijing National Laboratory for Condensed Matter Physics, and Institute of Physics, Chinese Academy of Sciences, Beijing 100190, China}
\author{T.-T.~Zhang}\thanks{These authors contributed equally to this work}
\affiliation{Beijing National Laboratory for Condensed Matter Physics, and Institute of Physics, Chinese Academy of Sciences, Beijing 100190, China}
\author{C.-J.~Yi}
\affiliation{Beijing National Laboratory for Condensed Matter Physics, and Institute of Physics, Chinese Academy of Sciences, Beijing 100190, China}
\author{Z.-D.~Song}
\affiliation{Beijing National Laboratory for Condensed Matter Physics, and Institute of Physics, Chinese Academy of Sciences, Beijing 100190, China}
\author{W.-L.~Zhang}
\affiliation{Department of Physics $\&$ Astronomy, Rutgers University, Piscataway, New Jersey 08854, USA}
\author{T.~Zhang}
\affiliation{Beijing National Laboratory for Condensed Matter Physics, and Institute of Physics, Chinese Academy of Sciences, Beijing 100190, China}
\author{Y.-G.~Shi}
\affiliation{Beijing National Laboratory for Condensed Matter Physics, and Institute of Physics, Chinese Academy of Sciences, Beijing 100190, China}
\author{H.-M.~Weng}
\affiliation{Beijing National Laboratory for Condensed Matter Physics, and Institute of Physics, Chinese Academy of Sciences, Beijing 100190, China}
\affiliation{Collaborative Innovation Center of Quantum Matter, Beijing, China}
\author{Z.~Fang}
\affiliation{Beijing National Laboratory for Condensed Matter Physics, and Institute of Physics, Chinese Academy of Sciences, Beijing 100190, China}
\affiliation{Collaborative Innovation Center of Quantum Matter, Beijing, China}
\author{P.~Richard}\email{p.richard@iphy.ac.cn}
\affiliation{Beijing National Laboratory for Condensed Matter Physics, and Institute of Physics, Chinese Academy of Sciences, Beijing 100190, China}
\affiliation{Collaborative Innovation Center of Quantum Matter, Beijing, China}
\affiliation{School of Physical Sciences, University of Chinese Academy of Sciences, Beijing 100190, China}
\author{H.~Ding}\email{dingh@iphy.ac.cn}
\affiliation{Beijing National Laboratory for Condensed Matter Physics, and Institute of Physics, Chinese Academy of Sciences, Beijing 100190, China}
\affiliation{Collaborative Innovation Center of Quantum Matter, Beijing, China}
\affiliation{School of Physical Sciences, University of Chinese Academy of Sciences, Beijing 100190, China}

\date{\today}

\begin{abstract}
We report a polarized Raman scattering study of non-symmorphic topological insulator KHgSb with hourglass-like electronic dispersion. Supported by theoretical calculations, we show that the lattice of the previously assigned space group $P6_3/mmc$ (No. 194) is unstable in KHgSb. While we observe one of two calculated Raman active E$_{2g}$ phonons of space group $P6_3/mmc$ at room temperature, an additional A$_{1g}$ peak appears at 99.5 ~cm$^{-1}$ upon cooling below $T^*$ = 150 K, which suggests a lattice distortion. Several weak peaks associated with two-phonon excitations emerge with this lattice instability. We also show that the sample is very sensitive to high temperature and high laser power, conditions under which it quickly decomposes, leading to the formation of Sb. Our first-principles calculations indicate that space group $P6_3mc$ (No. 186), corresponding to a vertical displacement of the Sb atoms with respect to the Hg atoms that breaks the inversion symmetry, is lower in energy than the presumed $P6_3/mmc$ structure and preserves the glide plane symmetry necessary to the formation of hourglass fermions.
\end{abstract}

\pacs{}

\maketitle

\section{Introduction}

On account of emergent phenomena, the concepts of high-energy physics can be realized in condensed matter physics. For example, Dirac fermions are found in the form of Dirac cones in low-energy excitations in graphene, topological insulators \cite{graphene,TI1} and even in the parent compounds of Fe-based superconductors \cite{Ran_PRB79,RichardPRL2010}. Majorana fermions exist as quasiparticle excitations in superconductors \cite{Majorana1,Majorana2}, while Weyl fermions have been predicted \cite{X_Wan_PRB83,Balents_Physics2011,G_Xu_PRL107,weyl2,Huang_NCOMM} and detected by angle-resolved photoemission spectroscopy (ARPES) in Weyl semimetals \cite{weyl1,SY_Xu_Science2015,Yang_LX_NPHYS2015,Lv_BQ_nphys11}. Due to looser symmetry restrictions, condensed matter systems constitute a vast playground for discovering exotic fermions that have no equivalent in high-energy physics \cite{newfermion}.

Recently, an exotic hourglass electronic dispersion has been predicted \cite{hourglass_nature} at the surface of topological insulator KHgSb, and an electronic structure consistent with the calculations has been observed by ARPES \cite{hourglass_ARPES}. Due to limited energy resolution in the ARPES measurements, the existence of the hourglass surface states relies for a large part on the calculations based on the space group $P6_3/mmc$ (No. 194) of the crystal structure determined at room temperature for this material \cite{falsestructureinTI}. However, this material is difficult to grow and to handle due to its air-sensitivity, and literature lacks of experimental characterization of the lattice properties of KHgSb at low temperature, where the exotic hourglass state is observed. Whether the crystal structure of KHgSb is preserved at low temperature, is unclear.

In this paper, we report a structure instability in KHgSb using both Raman scattering and first-principles calculations. At room temperature, we observe one optical mode at 143.8 ~cm$^{-1}$, with energy and symmetry consistent with one of the two calculated Raman active E$_{2g}$ modes. An additional hump with A$_{1g}$ symmetry is detected at 87~cm$^{-1}$. As temperature decreases, a new peak with A$_{1g}$ symmetry shows up at 99.5 ~cm$^{-1}$ below $T^*$ = 150 K. This violation of the Raman selection rules indicates an imperfect arrangement of the lattice. Consistently, our first-principles calculations of KHgSb with space group $P6_3/mmc$ (No. 194) show a virtual A$_{2u}$ mode at the Brillouin zone (BZ) center ($\Gamma$). Based on the vibrations of this virtual mode involving Hg an Sb displacements along the Z axis, we find that the structure characterized by subgroup ($P6_3mc$, No. 186), which we obtain by displacing the Hg atoms and Sb atoms a little in opposite directions along the Z axis, has a similar total energy as the undistorted structure. Our first-principles calculations in group $P6_3mc$ show a A$_1$ Raman active mode at 96.2~$^{-1}$, which is in good agreement with the observed A1g mode at 99.5~$^{-1}$ below $T^*$. Upon further cooling, the hump around 87 ~cm$^{-1}$ splits into two components, and new peaks enhanced by the lattice distortion emerge at higher energy, which we assign to multiple-phonon excitations based on calculations of the combined phonon density-of-states. We also show that the sample decomposes quickly into Sb above 320 K or under high laser power, reinforcing the claim of structure instability. Our theoretical analysis indicates that the existence of the hourglass states at the surface of KHgSb is robust to the structural instability observed by Raman scattering.

The current manuscript is structured as follows. We describe the technical experimental setup and the computations methods in Section \ref{section_Experiment} and Section \ref{section_Computation}, respectively. In Section \ref{section_evidences}, we present experimental and computational evidences for a lattice instability of KHgSb with space group No. 194, and we characterize a lattice distortion occurring below $T^*=$~150~K. In Section \ref{robustness_check}, before the summary, we present calculations of the surface states of the distorted structure and check the robustness of the hourglass dispersion upon distortion.

\begin{figure*}[!t]
\begin{center}
\includegraphics[width=\textwidth]{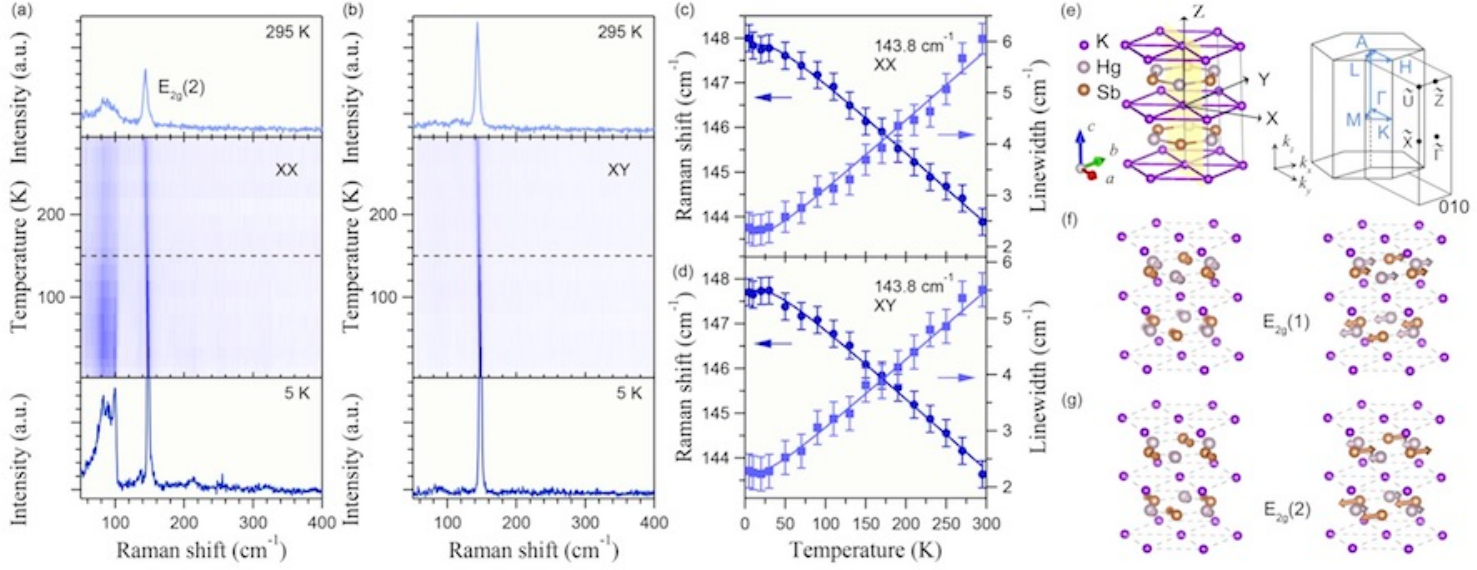}
\end{center}
\caption{\label{figure1}(Color online). (a) Intensity plot of the Raman shift as a function of temperature under the (XX) polarization configuration. The upper and lower panels show the Raman spectra recorded at 295 K and 5 K, respectively. (b) Same as (a) but under the (XY) polarization configuration. The dashed lines in (a) and (b) indicate $T^*$ = 150 K. (c) Temperature dependence of the Raman shift (left axis) and linewidth (right axis) of the E$_{2g}$(2) mode at 143.8 ~cm$^{-1}$ under the (XX) polarization configuration. The lines are fitting curves according to the standard anharmonic-decay model \cite{formular2}. (d) Same as (c) but under the (XY) polarization configuration. (e) Crystal structure (space group No. 194) and first BZ denoting high-symmetry points in the momentum space. The gray parallelepiped shows the unit cell and the yellow plane shows one of the glide-mirrors. The colored coordinates and the black coordinates are the unit cell axes and definition of directions used in this article, respectively. (f)-(g) Atomic displacements for the two E$_{2g}$ Raman active modes. The displacements of the atoms are indicated by arrows. }
\end{figure*}

\section{Experiment}\label{section_Experiment}

The single crystals of KHgSb used in our experiments have been grown using the self-flux method \cite{hourglass_ARPES}. Because they are highly air-sensitive, the samples were prepared in a glove box and cleaved before measurements to obtain clean surfaces. Temperature dependence studies between 5 and 430 K were carried out in a cryostat with a working vacuum better than $2\times 10^{-6}$ mbar. Back-scattering micro-Raman measurements were performed with the 514.5 nm and 488.0 nm excitations of an Ar-Kr laser focused on good sample surface regions by a $20\times$ objective. The power at the sample was kept smaller than 0.1 mW to avoid laser heating, except for the decomposition tests using up to 0.62 mW. The Raman scattering signal was analyzed by a Horiba Jobin Yvon T64000 spectrometer equipped with a nitrogen cooled CCD camera.

The lattice vibrations can be classified in terms of the irreducible representations of the point group characterizing the crystal studied. This symmetry classification defines how these excitations transform under the symmetry operations of the crystal group. By controlling the incident ($\mathbf{\hat{e}}^i$) and scattered ($\mathbf{\hat{e}}^s$) polarization wave vectors in the experimental geometry ($\mathbf{\hat{e}}^i\mathbf{\hat{e}}^s$), it is possible to separate the excitations of distinct symmetries based on Raman selection rules. As shown in Fig. \ref{figure1}(e), we define X as the direction along the Hg-Sb bonds, and Y as the direction of the unit cell vector \emph{b}, which is within the glide-mirrors. The Z direction corresponds to the axis perpendicular to the HgSb planes. Raman spectra are recorded under the ($\mathbf{\hat{e}}^i\mathbf{\hat{e}}^s$) = (XX) and (XY) polarization configurations.

\section{Computational methods}\label{section_Computation}

We performed first-principles calculations to obtain the phonon dispersions and the stable subgroups of KHgSb in the framework of the density functional perturbation theory (DFPT) \cite{DFPT2}. The experimental lattice constants and fully-relaxed data are shown in Table \ref{lattice_constant}. For both electronic and phonon calculations, we used the Vienna \textit{ab-initio} simulation package (VASP) \cite{VASP1} with the generalized gradient approximation (GGA) of Perdew-Burke-Ernzerhof for the exchange-correlation functions \cite{GGA}. The projector augmented wave (PAW) \cite{PAW} method was employed to describe the electron-ion interactions. A plane wave cut-off energy of 500 eV was used with a uniform 10$\times$10$\times$5 Monkhorst-Pack $k$-point mesh for integrations over the Brillouin zone for a 3$\times$3$\times$1 supercell. The frequencies and displacement patterns of the phonon modes were derived from the dynamical matrix generated by the DFPT method. The Wannier center for space group No. 186 is also obtained by using VASP.

For characterizing the topological nature of space group No. 186, we calculated the Berry phase for all occupied electronic bands by using the Wilson loop method \cite{WL1,WL2}. We calculated the Berry phase of band $n$ from a 1D integration of Berry connection along the $k_y$ axis: $\theta_n$($k_y$) = $\int_{-\pi}^{\pi}A_{n,\perp}dk_{\perp}$. Here $k_{\perp}$ is the reciprocal lattice vector on the surface that is perpendicular to $k_y$, while $A_{n,\perp}$ is the Berry connection of band $n$. As discussed in Ref. \cite{WL3}, $\theta_n$($k_y$) can be regarded as the 1D Wannier center of band $n$ and the Chern number can be described by the winding number of $\theta_n$($k_y$) when $k_y$ evolves over a periodic cycle. The Chern number can also be described by the number of edge states for topological insulators with a boundary, and thus the shape of the Wilson loop is isomorphic to the surface states, as shown in Fig. \ref{Sfigure1}.

\begin{table}
\caption{\label{lattice_constant}Relaxed lattice constants used in calculations.}
\begin{ruledtabular}
\begin{tabular}{cccc}
  &	No. 194   & No. 194  &	 No. 186 \\
  & Experiment \cite{sample_grow} & Relaxed &  Relaxed\\
\hline
\emph{a}& 4.784 &  4.892 & 4.892 \\
\emph{b}& 4.784 &  4.892 & 4.892 \\
\emph{c}& 10.225 & 10.516 & 10.527 \\
\end{tabular}
\end{ruledtabular}
\begin{raggedright}
The values are given in \AA.\\
\end{raggedright}
\end{table}

\section{Evidences of lattice instability} \label{section_evidences}
\subsection{Lattice distortion at low temperature}

An earlier study reported that the crystal structure of KHgSb at room temperature is characterized by space group $P6_3/mmc$ (point group $D_{6h}$) \cite{sample_grow}. A simple group symmetry analysis \cite{bilbal} indicates that the phonon modes at the $\Gamma$ point decompose into [2E$_{2g}$+2A$_{2u}$+2E$_{1u}$+2B$_{1g}$+B$_{2u}$+E$_{2u}$]+[A$_{2u}$+E$_{1u}$], where the first and second terms represent the optic modes and the acoustic modes, respectively. Among the optic modes, only the two E$_{2g}$ modes are Raman active, while the A$_{2u}$ and E$_{1u}$ modes are infrared active. The other modes are silent. As shown in Figs. \ref{figure1}(f) and \ref{figure1}(g), the corresponding atomic vibrations of the two Raman active E$_{2g}$ modes involve the Hg and Sb motions within the HgSb planes. The Raman tensors corresponding to the $D_{6h}$ symmetry group are expressed in the XYZ coordinates as (note that the A$_{1g}$ and E$_{1g}$ modes are absent in KHgSb):
\begin{displaymath}
\textrm{A$_{1g}=$}
\left(\begin{array}{ccc}
a & 0 &0\\
0 & a &0\\
0 & 0 &b
\end{array}\right),
\end{displaymath}
\begin{displaymath}
\left[\begin{array}{ccc}\textrm{E$_{1g}=$}
\left(\begin{array}{ccc}
0 & 0 &0\\
0 & 0 &c\\
0 & c &0\\
\end{array}\right),
\left(\begin{array}{ccc}
0 & 0 &-c\\
0 & 0 &0\\
-c & 0 &0\\
\end{array}\right)
\end{array}\right],
\end{displaymath}
\begin{displaymath}
\left[\begin{array}{ccc}\textrm{E$_{2g}=$}
\left(\begin{array}{ccc}
d & 0 &0\\
0 & -d &0\\
0 & 0 &0\\
\end{array}\right)
, \left(\begin{array}{ccc}
0 & -d &0\\
-d & 0 &0\\
0 & 0 &0\\
\end{array}\right)
\end{array}\right].
\end{displaymath}
\noindent The temperature evolution of the Raman spectra in the $(\mathbf{\hat{e}}^i\mathbf{\hat{e}}^s)$= (XX) and (XY) polarization configurations are shown in Figs. \ref{figure1}(a) and \ref{figure1}(b), respectively. Based on polarization selection rules, the modes appearing in both of these two channels have the E$_{2g}$ symmetry. This is the case of the peak at 143.8 ~cm$^{-1}$, which is obviously one of the two E$_{2g}$ modes of KHgSb. The temperature dependence of its energy and linewidth are analyzed in Figs. \ref{figure1}(c) and \ref{figure1}(d), respectively. The results are in line with the standard anharmonic-decay model \cite{formular2}.

\begin{figure}[!t]
\begin{center}
\includegraphics[width=\columnwidth]{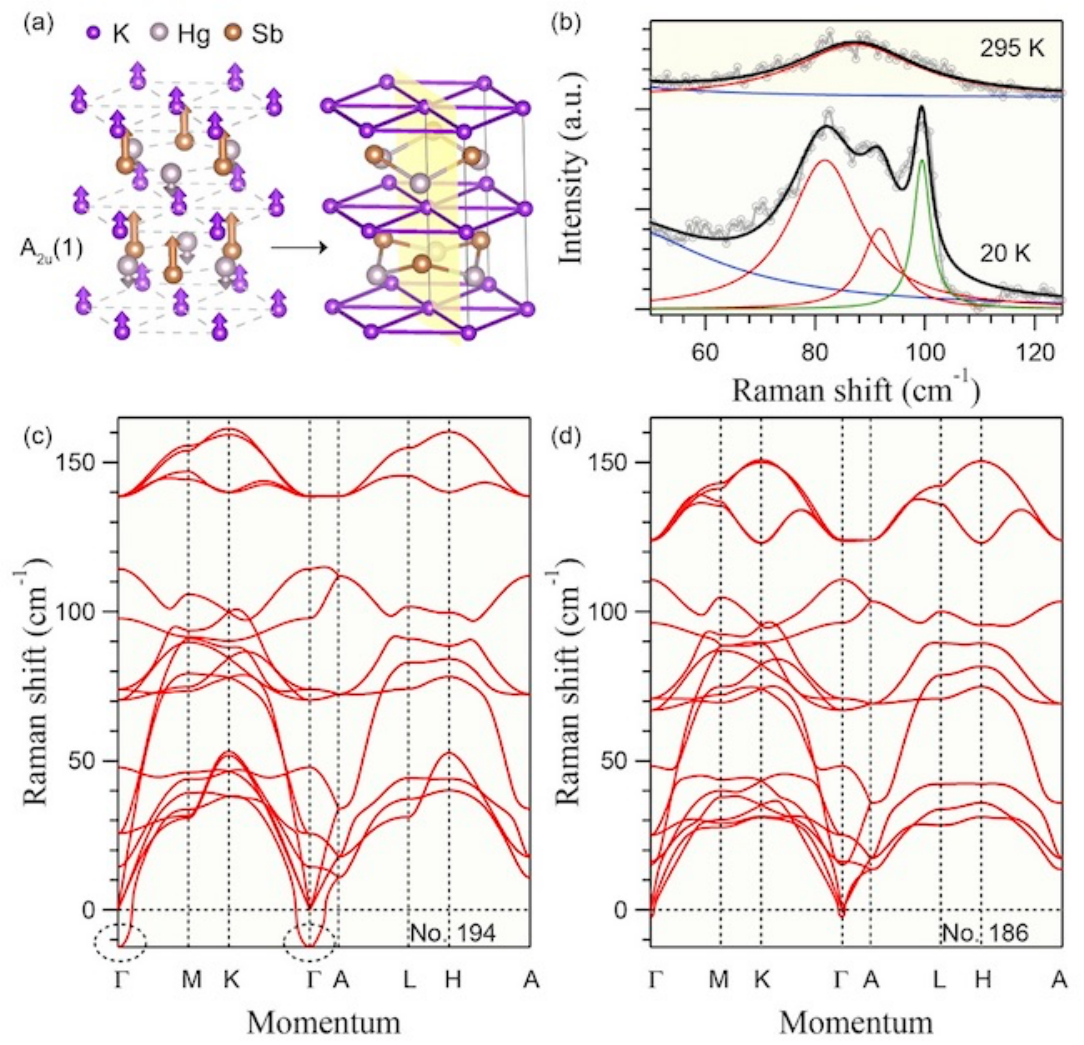}
\end{center}
\caption{\label{figure2}(Color online). (a) Atomic displacements for the virtual optic A$_{2u}$ mode and resulting crystal structure with space group No. 186. The distance between Hg and Sb atoms are exaggerated for better visualisation. The gray parallelepiped shows the unit cell and the yellow plane shows one of the glide-mirrors. (b) Comparison of spectra measured at 295~K and 20~K in the range of the hump with A$_{1g}$ symmetry. Individual Lorentz peak functions are represented by red (two-phonon excitations) and green curves (one-phonon excitation) while the blue curves are background functions. The bolder black curves represent the resulting fitted spectra. (c) Calculation of the phonon band dispersion with space group No.~194 along high-symmetry lines. The imaginary frequencies are shown as negative and indicated by dashed ellipses. The momentum path is illustrated by blue arrows in the right panel of Figs. \ref{figure1}(e). (d) Same as (c) but with space group No.~186. }
\end{figure}

Interestingly, we observe an additional hump at 87~cm$^{-1}$ with the A$_{1g}$ symmetry (it appears only under the parallel polarizations configuration), for which no Raman active mode is expected. As temperature decreases, a new peak with A$_{1g}$ symmetry appears at 99.5~cm$^{-1}$ below $T^*$~=~150~K. Since no A$_{1g}$ mode is predicted for KHgSb with space group No.~194, this new A$_{1g}$ peak implies either a distortion of the lattice breaking the selection rules or an impurity phase. With further cooling, the hump spits into two components, and new peaks with weaker intensity emerge.

\begin{table*}
\caption{\label{CAL}Comparison of the calculated optical phonon modes of space groups No. 194 ($D_{6h}$) and No. 186 ($C_{6v}$), along with the experimental results.}
\footnotesize\rm
\begin{ruledtabular}
\begin{tabular}{cccccccccccc}
No. 194& \textcolor{red}{A$_{2u}$(1)}& B$_{1g}$& \textbf{E$_{2g}$(1)}& B$_{1g}$& E$_{2u}$& E$_{1u}$(1)& A$_{2u}$(2)& B$_{2u}$& E$_{1u}$(2)& \textbf{E$_{2g}$(2)}\\
Cal.& \textcolor[rgb]{1.00,0.00,0.00}{-12.4}& 14.5& 25.6& 47.8& 70.4& 74.0& 97.7& 114.4& 138.5& 138.7\\
\hline
No. 186& \textbf{A$_{1}$(1)}& B$_{1}$(1)& \textbf{E$_{2}$(1)}& B$_{1}$(2)& \textbf{E$_{2}$(2)}& \textbf{E$_{1}$(1)}& \textbf{A$_{1}$(2)}& B$_{1}$(3)& \textbf{E$_{1}$(2)}& \textbf{E$_{2}$(3)}\\
Cal.& 15.2& 16.3& 25.1& 48.2& 67.0& 70.9& 96.2& 110.7& 123.7& 124.1\\
\hline
Exp.& & & & & & & 99.5 (5 K)& & & 143.8 (295 K)\\
\end{tabular}
\begin{raggedright}The peak positions are given in ``cm$^{-1}$". The virtual mode is marked in red. The Raman active modes are emphasized in bold.\\
\end{raggedright}
\end{ruledtabular}
\end{table*}

Our first-principles calculations also suggest the instability of the crystal structure of KHgSb based on space group No. 194. We show in Fig. \ref{figure2}(c) the corresponding phonon dispersion calculations, and the mode energies at the $\Gamma$ point are given in Table \ref{CAL}, along with their symmetries. Although the experimentally-observed peak at 143.8 ~cm$^{-1}$ is in accord with the calculated E$_{2g}$(2) mode, the calculations exhibit a A$_{2u}$  virtual mode (negative energy), which means that the restoring force is not large enough to pull the atoms back into their equilibrium positions. In such circumstance, the crystal structure is intrinsically prone to structural distortions.

\begin{figure}[!t]
\begin{center}
\includegraphics[width=\columnwidth]{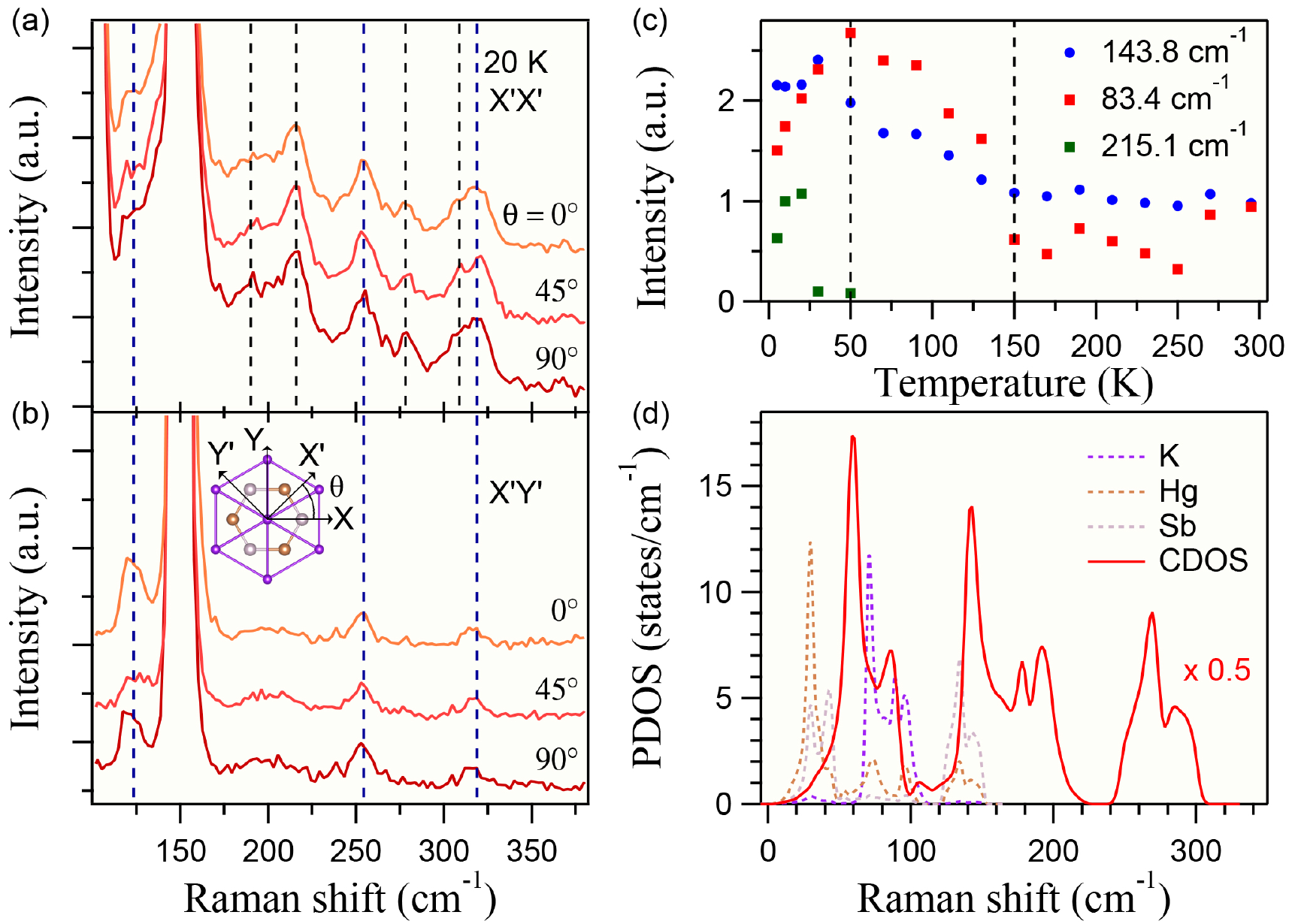}
\end{center}
\caption{\label{figure3}(Color online). (a) Second-order Raman spectra recorded at 20 K along different directions under parallel polarization configuration and (b) perpendicular polarization configuration. The inset in (b) illustrates the definition of $\theta$, which is the angle between X' (Y') and X (Y). The black and blue dashed lines indicate two-phonon peaks with A$_1$ and E$_2$ symmetries, respectively. (c) Temperature dependence of the integrated intensity under the (XX) polarization configuration of the E$_{2g}$(2) mode at 143.8~cm$^{-1}$ (blue dots), as well as the two-phonon excitations at 83.4~cm$^{-1}$ (red squares) and 215.1~cm$^{-1}$ (green squares). The dashed lines indicate $T$ = 50 K and $T^*=150$ K. The data have been rescaled for better visualization. (d) Phonon density-of-states for each elements (dashed lines) and combined density-of-states (CDOS, represented by solid red lines) of KHgSb with the space group No.~186 structure.}
\end{figure}

A possible stable structure can be guessed according to the atomic vibrations of this A$_{2u}$ virtual mode. As illustrated in Fig. \ref{figure2}(a), this A$_{2u}$ optic mode is mainly related to opposite vibrations of Hg and Sb atoms along the Z axis. By moving the Sb atoms a little higher than the Hg atoms, we obtain a new structure characterized by space group No. 186 (point group $C_{6v}$), which is a subgroup of space group No. 194. In this new structure, the inversion symmetry is broken. Among all 7 possible subgroups having the same unit cell as the original structure, only space group No. 186 is stable according to our calculations.

\begin{figure*}[!t]
\begin{center}
\includegraphics[width=\textwidth]{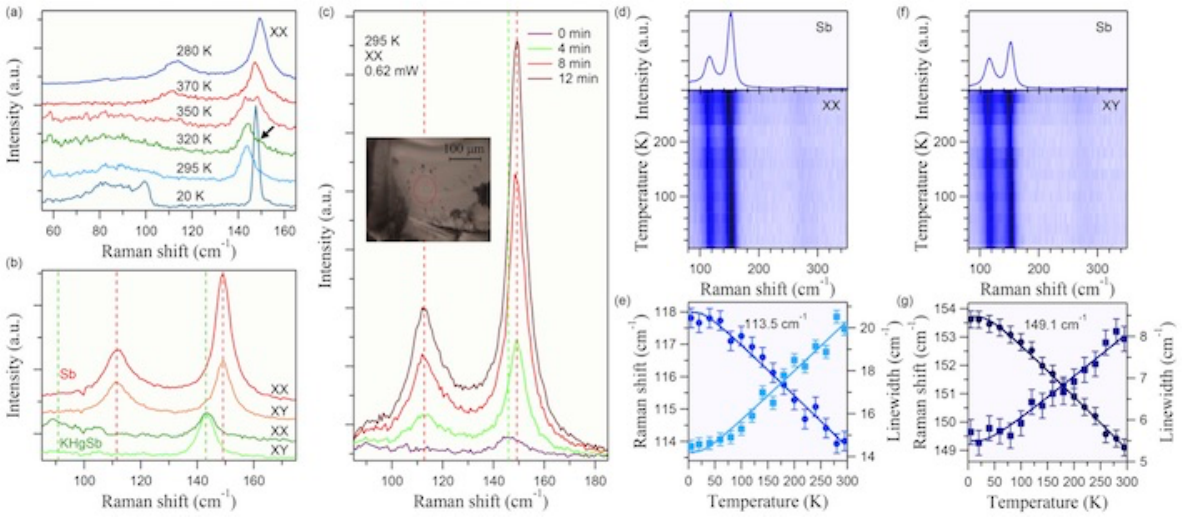}
\end{center}
\caption{\label{figure4}(Color online). (a) Raman spectra in the (XX) channel at temperature varying from 20~K to 370~K, and then back to 280~K. The arrow indicates the beginning of decomposition. (b) Comparison of Raman spectra from KHgSb and Sb under different polarization configurations. The green and red dashed lines indicate the Raman signals of KHgSb and Sb, respectively. (c) Time evolution of the Raman spectra in the (XX) channel at room temperature and under stronger laser intensity (0.62~mW). The inset shows the laser spots changing to yellow color because of the decomposition. (d)-(g) Raman results for Sb (space group $R\overline{3}m$ (No. 166)): (d) Intensity plot of the Raman shift as a function of temperature under the (XX) polarization configuration. The upper panel shows the Raman spectra recorded at 295~K. (e) Temperature dependence of the Raman shift (left axis) and linewidth (right axis) of E$_{g}$ mode at 113.5~cm$^{-1}$ under the (XX) polarization configuration. The curves are fitting equations according to standard anharmonic-decay model \cite{formular2}. (f) Same as (d) but under the (XY) polarization configuration. (g) Same as (e) but for the A$_{1g}$ mode at 149.1~cm$^{-1}$.}
\end{figure*}

The calculations of the lattice dynamics of KHgSb based on space group No. 186 are presented in Fig. \ref{figure2}(d) and Table \ref{CAL}. No virtual mode is obtained. The phonon modes at the $\Gamma$ point decompose into [2A$_1$+2E$_1$+3E$_2$+3B$_1$]+[A$_1$+E$_1$], where the first and second terms represent the optic modes and the acoustic modes, respectively. Among the optic modes, A$_1$ and E$_1$ are both Raman and infrared active, E$_2$ is Raman active, and B$_1$ is silent. The Raman tensors related to the corresponding $C_{6v}$ point group symmetry are expressed in the XYZ coordinates as:
\begin{displaymath}
\textrm{A$_{1}=$}
\left(\begin{array}{ccc}
a & 0 &0\\
0 & a &0\\
0 & 0 &b
\end{array}\right),
\end{displaymath}
\begin{displaymath}
\left[\begin{array}{ccc}\textrm{E$_{1}(x)=$}
\left(\begin{array}{ccc}
0 & 0 &c\\
0 & 0 &0\\
c & 0 &0\\
\end{array}\right),\textrm{E$_{1}(y)=$}
\left(\begin{array}{ccc}
0 & 0 &0\\
0 & 0 &c\\
0 & c &0\\
\end{array}\right)
\end{array}\right],
\end{displaymath}
\begin{displaymath}
\left[\begin{array}{ccc}\textrm{E$_{2}=$}
\left(\begin{array}{ccc}
d & 0 &0\\
0 & -d &0\\
0 & 0 &0\\
\end{array}\right)
, \left(\begin{array}{ccc}
0 & -d &0\\
-d & 0 &0\\
0 & 0 &0\\
\end{array}\right)
\end{array}\right].
\end{displaymath}

We note that the E$_1$ symmetry modes cannot be detected here because the incident and scattered light are along the Z axis in our study. As shown in Figs. \ref{figure1}(a) and \ref{figure2}(b), the new peak showing up below $T^*$ = 150 K is consistent with the A$_{1}$(2) mode in the space group No.~186 calculations. The original E$_{2g}$(2) mode at 143.8~cm$^{-1}$ corresponds to the E$_2$(3) mode in the space group No.~186 representation. Although its integral intensity is enhanced below $T^*$~=~150~K, as shown in Figs. \ref{figure3}(c), no obvious anomaly is found in the peak position or linewidth of this mode. Our first-principles calculations also give very similar energies for the two structures. The energy of KHgSb based on space group No.~186 is only 0.051~meV lower than the energy of space group No.~194. Thus the lattice instability observed in our experiments corresponds either to a rather small distortion in the HgSb plane or to a coexisting phase, which may produce disorder in the overall structure. As for the other calculated A$_{1}$ and E$_2$ Raman active modes, they are not observed because of their low energies, which are out of our observation range.

\subsection{Enhanced second-order Raman scattering}

Now we go back to the additional hump with A$_{1g}$ symmetry and the emerging peaks with weaker intensities below $T^*$. As shown in Fig. \ref{figure2}(b), the hump with energy around 87~cm$^{-1}$ at room temperature has two components, 83.4~cm$^{-1}$ and 90.1~cm$^{-1}$, with intensity enhanced at low temperature. These two peaks, as well as the new emerging peaks at 191.4~cm$^{-1}$, 215.1~cm$^{-1}$, 278.9~cm$^{-1}$ and 309.5~cm$^{-1}$ are observed only under parallel polarization configurations, while the peaks at 122.1~cm$^{-1}$, 253.0~cm$^{-1}$ and 316.6~cm$^{-1}$ are observed under both parallel and perpendicular polarization configurations, as shown in Figs. \ref{figure3}(a) and \ref{figure3}(b). The linewidths of these peaks are broader than the one-phonon peaks, and most of them have higher energies than the phonon energy dispersion range displayed in Fig. \ref{figure2}(d). Consequently, we ascribed them to multiple-phonon excitations.

For second-order Raman scattering, the wavevectors of the participating phonons need not to be near zero, but can take values throughout the Brillouin zone, as long as their total wavevector is zero ($Q\equiv q + q' = 0$). In a general case, the symmetries of the modes vary with momentum and different excitations can be combined. In particular, lowering the crystal symmetry relaxes the Raman selection rules for these excitations. Multiple-phonon excitations can thus be detected in both the (XX) and (XY) channels. The scattering intensity of scattering at Raman shift $\omega$ is controlled by the number of pairs of phonons with total frequency $\omega$ \cite{book}. In Fig. \ref{figure3}(d), we report the phonon density-of-states (PDOS) of each elements and the combined density-of-states (CDOS) for sums of two phonons with total wavevector $Q$ equal to zero (based on the space group No.~186 structure), which is consistent with our measurements. As shown in Fig. \ref{figure3}(c), the intensity of the two-phonon excitation at 83.4~cm$^{-1}$ increases just below $T^*$. With further cooling, more two-phonon excitations like the one at 215.1~cm$^{-1}$ emerge below 50 K. The second-order Raman scattering enhancing below $T^*$ reinforces our assumption that a lattice distortion occurs below that temperature.

\subsection{Decomposition at high temperature}

Then we report the behavior of KHgSb above room temperature. As shown in Fig. \ref{figure4}(a), the typical Raman peaks change around 320 K when increasing temperature from low temperature, and the spectra become completely different at 370~K. Since the spectrum does not recover when temperature decreases back below room temperature, this change is associated with a deterioration rather than to a phase transition. We show in Fig. \ref{figure4}(c) that similar deterioration is found in the laser heating tests we made with a higher laser power (0.62~mW). The spectrum changes with laser exposition time. The sample area at the laser spot changes gradually from gray to yellow. The temperature evolution of the Raman spectra after deterioration of KHgSb are shown in Figs. \ref{figure4} (d)-(g). By comparison with earlier work in the literature, we identify the two peaks at 113.5~cm$^{-1}$ and 149.1~cm$^{-1}$ observed at the decomposed yellow area to characteristic Raman peaks of Sb \cite{Sb_Raman_1,Sb_Raman_2}. The hump at around 270~cm$^{-1}$ corresponds to second-order Raman signal \cite{Sb_Raman_1}.

The Raman spectra of KHgSb and Sb are compared in Fig. \ref{figure4}(b). The spectrum of Sb has two Raman active modes, and can easily be mistaken for that of KHgSb. However, unlike pristine KHgSb, the second-order Raman signal in Sb can be observed from 5~K to 295~K, with no anomalous change happening with decreasing temperature. Since we do not detect any additional trace of impurity, notably involving Hg or K, the precise decomposition process remains unclear. We note that this decomposition here occurs in vacuum, and no peak is detected by Raman scattering for samples exposed to air. In any case, the decomposition of KHgSb at relatively low temperature or under relatively moderate laser power further supports our assumption that the group No. 194 structure of KHgSb is unstable and prone to distortion.

\section{Robustness of hourglass electronic dispersion}\label{robustness_check}

\begin{figure}[!t]
\begin{center}
\includegraphics[width=\columnwidth]{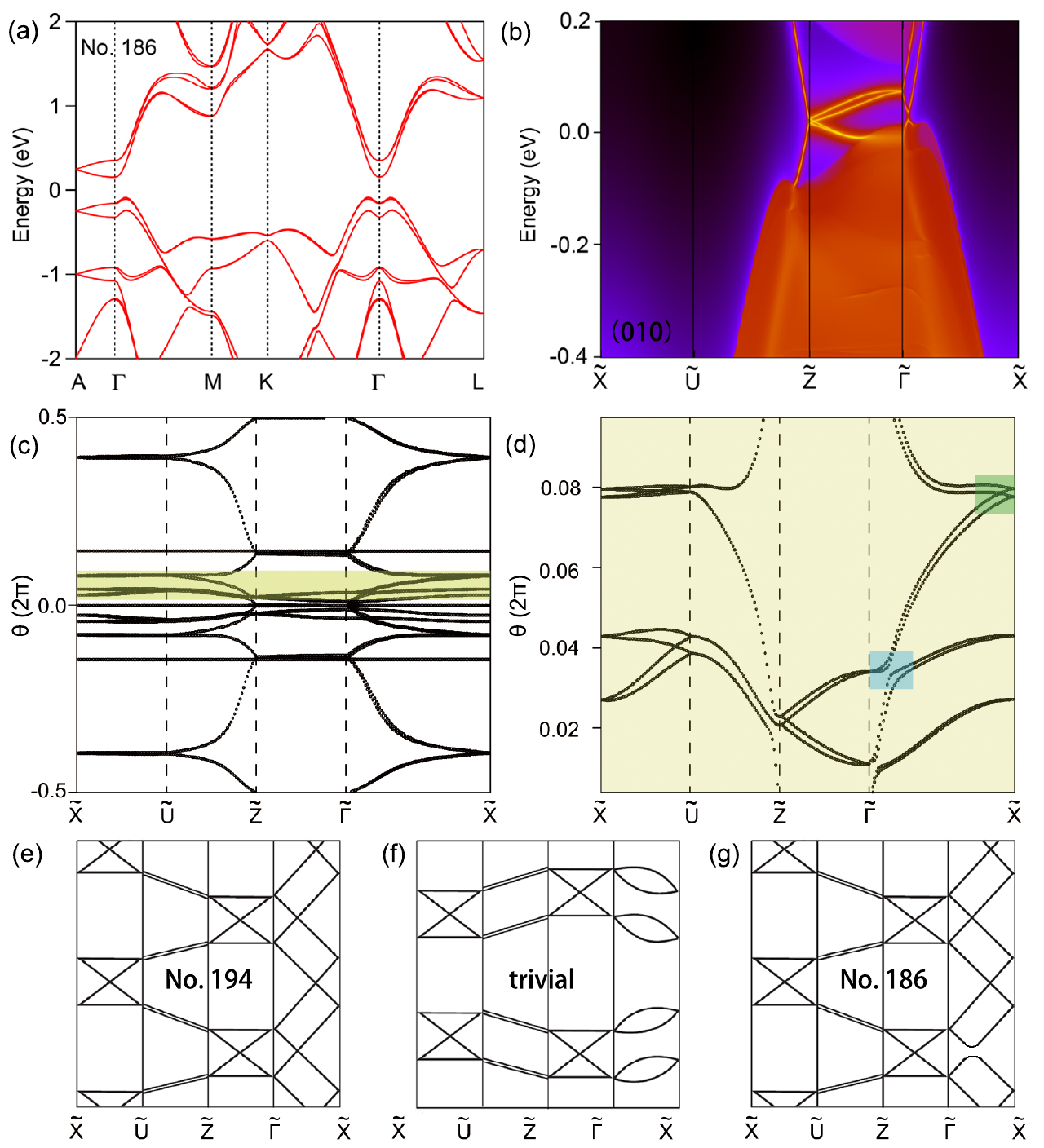}
\end{center}
\caption{\label{Sfigure1}(Color online). (a) Bulk band structure and (b) (010) surface states of KHgSb with space group No. 186. (c) Average position of the Wannier centers obtained by Wilson-loop calculation of KHgSb with space group No.~186. (d) Close-up of the light green region in (c). The dark green part and the blue part show the band crossing and the gap, respectively. (e) Non-trivial (KHgSb with space group No. 194) and (f) trivial topologies of hourglass surface states illustrated in Ref. \cite{hourglass_nature}. (g) (010) surface topologies of KHgSb with space group No. 186. }
\end{figure}

In Section \ref{section_evidences} we showed several experimental and theoretical evidences for a lattice instability in KHgSb. We now ask: is the hourglass dispersion predicted for KHgSb with space group No. 194 robust to the lattice instability detected by Raman scattering? In order to answer this question, we performed first-principles calculations of the distorted structure. The bulk band structure of distorted KHgSb with space group No. 186 is illustrated in Fig. \ref{Sfigure1} (a). The bulk electronic structure calculated for the distorted structure is very similar to that reported in Ref. \cite{hourglass_nature} for the undistorted one, with the exception of the lifting of the degeneracy for some degenerate bands due to the breakdown of the inversion symmetry after distortion that is induced by the fact that the Hg and Sb atoms are no longer located in the same plane. The similarity of the band structures of KHgSb with the space groups Nos. 194 and 186 explains why the electronic band structure of KHgSb measured by ARPES is consistent with the band calculations of the undistorted structure \cite{hourglass_ARPES}.

What about the surface states? Our surface state calculations on the (010) surface are displayed in Fig. \ref{Sfigure1} (b). Interestingly, we still found hourglass electronic dispersion on this surface along $\widetilde{\Gamma}\widetilde{Z}$ and $\widetilde{X}\widetilde{U}$ in group No. 186. Again, this means that the structure proposed in this work, characterized by space group No. 186, is consistent with the recent ARPES results \cite{hourglass_ARPES}. The physical origin of the robustness of the hourglass dispersion lies in the preservation, despite a lattice distorsion breaking the inversion symmetry, of the non-symmorphic glide-mirror $\overline{M_x}$ symmetry, which along with time-reversal symmetry protects the topology of the hourglass surface states \cite{hourglass_nature,Cohomology_prx}. To confirm the non-trivial nature of this hourglass state, we applied the Wilson-loop method to KHgSb with the space group No. 186 structure. Figs. \ref{Sfigure1}(c) and \ref{Sfigure1}(d) show the corresponding Wilson-loop calculation results, and schematic illustrations of the non-trivial (KHgSb with space group No. 194) and trivial connections of hourglass states proposed by Zhijun Wang \emph{et al.} \cite{hourglass_nature}, as well as the case of distorted KHgSb with space group No. 186, are displayed in Figs. \ref{Sfigure1}(e)-\ref{Sfigure1}(g). Although the surface state bands can hybridize and open a gap along $\widetilde{\Gamma}\widetilde{X}$ due to the absence of the $\overline{M_z}$ symmetry operation in space group No. 186, the zigzag connection along $\widetilde{U}\widetilde{Z}$ indicates a non-trivial property. In other words, KHgSb remains a non-trivial topological insulator at low temperature and yes, the hourglass dispersion predicted for KHgSb with space group No. 194 is robust to the lattice instability detected by Raman scattering.

\section{Summary}

In summary, we performed a polarized Raman scattering and first-principles calculations of the lattice dynamics study of the non-symmorphic topological insulator KHgSb predicted to have hourglass topological state. At room temperature, we observed one of the two calculated Raman active E$_{2g}$ phonons based on the previously reported space group $P6_3/mmc$ (No. 194). At low temperature, an additional A$_{1g}$ peak inconsistent with the calculations emerges below $T^*$ = 150 K. With help of first-principles calculations, we identify the corresponding lattice distortion, which can be illustrated by a buckling lattice distortion of the HgSb layers with space group $P6_3mc$ (No. 186). This lattice distortion is accompanied by an enhancement of multiple-phonon excitations. The quick decomposition of samples upon increasing temperature from room temperature further indicates that the structure is unstable. Finally, our calculations show that dispersion has only effect on the electronic structure and that the topological hourglass surface state in KHgSb is preserved. Our work not only confirms the existence of a new exotic hourglass state, but also determines the ingredients of its robustness to lattice instability.

\vspace{1cm}
\section*{Acknowledgement}

We acknowledge Zhijun Wang, Quansheng Wu, A. Alexandradinata, J.-Z. Ma and T. Qian for useful discussions. This work was supported by grants from the Ministry of Science and Technology (Nos. 2015CB921301, 2016YFA0401000, 2016YFA0300300, 2016YFA0300600 and 2013CB921700) and the National Natural Science Foundation (Nos. 11274362 and 11674371, 11622435, 11474340, 11422428, 11274367, 11474330, 11504117 and 11234014) from China, and the Chinese Academy of Sciences (No. XDB07000000).

\bibliography{citation}

\end{document}